\documentclass[journal]{IEEEtran}
\ifCLASSINFOpdf
\else
\fi
\usepackage{graphicx}
\usepackage{amsthm}
\usepackage{amsmath,amssymb,amsfonts}
\usepackage{lineno}
\usepackage{balance}
\usepackage{algcompatible}
\usepackage{algorithm}
\usepackage{subcaption}
\usepackage{caption}
\usepackage[T1]{fontenc}
\usepackage[utf8]{inputenc}

\usepackage{optidef}
\usepackage{fixltx2e}
\usepackage{multirow}
\usepackage{xcolor}
\usepackage{textgreek}
\newtheorem{lemma}{Lemma}
\newtheorem{theorem}{Theorem}

\usepackage[acronym]{glossaries}
\makeglossaries
\newacronym{uav}{UAV}{Unmanned Aerial Vehicle}
\newacronym{uavs}{UAVs}{Unmanned Aerial Vehicles}
\newacronym{bs}{BS}{Base Station}
\newacronym{bss}{BSs}{Base Stations}
\newacronym{iot}{IoT}{Internet of Things}
\newacronym{qos}{QoS}{Quality of Service}
\newacronym{dbs}{DBS}{Drone Base Station}
\newacronym{dbss}{DBSs}{Drone Base Stations}
\newacronym{fbs}{FBS}{Flying Base Station}
\newacronym{fbss}{FBSs}{Flying Base Stations}
\newacronym{Mbps}{Mbps}{Megabit per second}
\newacronym{ga}{GA}{Genetic Algorithm}
\newacronym{los}{LoS}{Line of Sight}
\newacronym{nlos}{NLoS}{Non-Line of Sight}
\newacronym{fsm}{FSM}{FBS Set Management}
\newacronym{abs}{ABS}{Aerial Base Station}
\newacronym{abss}{ABSs}{Aerial Base Stations}
\newacronym{a2g}{A2G}{air-to-ground}
\newacronym{cso}{CSO}{cell switch-off}

\makeatletter

\begin{document}

\title{Managing Sets of Flying Base Stations \\ Using Energy Efficient $3$D Trajectory Planning \\ in Cellular Networks}


\author{\IEEEauthorblockN{
 Mohammad Javad Sobouti, Amir Hossein Mohajerzadeh, 
 Seyed Amin Hosseini Seno,
 and
 Halim Yanikomeroglu, \textit{Fellow, IEEE}
 }

\thanks{MohammadJavad Sobouti, Amir Hossein Mohajerzadeh, and Seyed Amin Hosseini Seno are with the Department of Computer Engineering, Ferdowsi University of Mashhad, Mashhad,Iran, (e-mail: javad.sobouti@mail.um.ac.ir; \{mohajerzadeh, hosseini\}@um.ac.ir).} 

\thanks{Halim Yanikomeroglu is with the Department of Systems and Computer Engineering, Carleton University, Ottawa, ON, Canada, (e-mail: halim@sce.carleton.ca).}
\thanks{The corresponding author is Amir Hossein Mohajerzadeh (email: mohajerzadeh@um.ac.ir).}
}

%



\maketitle

\begin{abstract}

Unmanned aerial vehicles (UAVs) in cellular networks have garnered considerable interest. One of their applications is as flying base stations (FBSs), which can increase coverage and quality of service (QoS). Because \acrshort{fbss} are battery-powered, regulating their energy usage is a vital aspect of their use; and therefore the appropriate placement and trajectories of FBSs throughout their operation are critical to overcoming this challenge. 
In this paper, we propose a method of solving a multi-\acrshort{fbs} $3$D trajectory problem that considers \acrshort{fbs} energy consumption, operation time, flight distance limits, and inter-cell interference constraints. Our method is divided into two phases: FBS placement and FBS trajectory. In taking this approach, we break the problem into several snapshots. First, we find the minimum number of \acrshort{fbss} required and their proper $3$D positions in each snapshot. Then, between every two snapshots, the trajectory phase is executed. The optimal path between the origin and destination of each \acrshort{fbs} is determined during the trajectory phase by utilizing a proposed binary linear problem (BLP) model that considers \acrshort{fbs} energy consumption and flight distance constraints. Then, the shortest path for each \acrshort{fbs} is determined while taking obstacles and collision avoidance into consideration. The number of FBSs needed may vary between snapshots, so we present an FBS set management (FSM) technique to manage the set of \acrshort{fbss} and their power. The results demonstrate that the proposed approach is applicable to real-world situations and that the outcomes are consistent with expectations.
\end{abstract}


\IEEEpeerreviewmaketitle

\section{Introduction} \label{intro}
\IEEEPARstart{C}{ellular} networks of the next generation offer increased data transfer rates, improved service quality, and greater energy efficiency. Cellular networks, including $5$G and beyond, are the backbone of future communication, capable of delivering a dependable and efficient infrastructure, resulting in a sustainable system.

Cellular networks have garnered substantial attention in recent decades and have significantly impacted human life. Due to the rapid growth in the popularity of cellular networks, communication vendors have been motivated to extend this beneficial technology and industry. For instance, the data transmission rate of cellular networks has increased dramatically from $1.2$ kbps in first-generation ($1$G) to $1$ Gbps or higher for fifth-generation ($5$G) and beyond networks.
As a result, $5$G is expected to be a greater step in the evolution of communication technology. Both $5$G and $6$G technologies will benefit businesses and society in the $2030$s by delivering highly reliable and secure communication services  \cite{foukas2017network, sheth2020taxonomy}.

Mobile user satisfaction in cellular networks depends on coverage and QoS metrics. As the quality of calls, movies, and the usage of various applications continues to increase, consumers want to enjoy this quality regardless of location, time, or circumstance, most notably on their smartphones. As a result, future generations of cellular networks' \acrshort{qos} and coverage must be improved. On the other hand, the increased frequencies required to serve a greater number of clients may result in coverage over a narrower region.
Therefore, additional base station (BS) are required to increase the coverage area, which results in a better quality of experience (QoE) for users. Increased terrestrial base station capacity is very costly, however, and almost unattainable. With the high financial requirements of creating a new base station, assessing its location and viability for installation is exceedingly complex and expensive. Additionally, increasing the bandwidth in future generations of the cellular network reduces the base station's coverage, assuming constant antenna power. As a result, fixing this matter will need more investigation. In the meantime, FBSs offer a viable way for expanding network coverage and \acrshort{qos} \cite{bekmezci2013flying, hayat2016survey}.

\acrshort{fbss} can thus be used when terrestrial base stations are uneconomical, or when they are unfeasible due to mountainous, rough, or rocky terrain. They can also be used when a cellular network is under heavy strain due to major sports or cultural events \cite{kalantari2017backhaul, alzenad20173}.
\acrshort{fbss} are beneficial since they do not require pre-arranged equipment and almost eliminate location limits. The other desirable feature of \acrshort{fbss} is their increased line-of-sight (LoS), which is due to their higher placement, thereby reducing multipath fading and shadowing \cite{fotouhi2019survey}.
Additionally, because \acrshort{fbss} are mobile, they can improve \acrshort{qos} and mitigate impairments by shifting their places if the network's status deteriorates. If necessary, a mobility function can also be used to enhance the number of users covered. Additionally, users may benefit from increased data rates by expanding the coverage and number of \acrshort{fbss} \cite{mozaffari2019tutorial}.

When using \acrshort{fbss} in cellular networks, their placements and altitudes must be specified. Since \acrshort{fbss} are often battery-powered, optimal positions and paths will extend the network's lifetime. Additionally, the locations of \acrshort{fbss} serving consumers are critical, as the major objective of cellular networks is to cover and deliver the level of service demanded by customers. However, because cellular users are mobile, they may move out of the range of a base station. To overcome this issue, users' positions are verified on a scheduled basis, and the location of \acrshort{fbss} may be identified using this information. After that, the \acrshort{fbss} are shifted.
While \acrshort{fbss} can circumvent the limitations of terrestrial base stations, they face location and trajectory challenges. Numerous studies have been conducted on deploying \acrshort{uav}s and stations in $2$D and $3$D space in various wireless networks \cite{sun2019location, zhang20213d}. Also, studies on the movement of \acrshort{fbss} involving wireless networks, the Internet of Things, and sensor networks have been undertaken \cite{khamidehi2021trajectory, shi2019multi}.

This paper considers cellular network demand in an urban area. We aim to cover users and serve their required data rate in a period of time using a $5$G and beyond cellular network. In doing so, the type of FBS we consider is the DJI S$900$ \acrshort{uav} helicopter, \cite{fotouhi2019survey} which can fly to an altitude of $3$ km\footnote{Regulations generally allow \acrshort{uav}s to fly up to about $120$ meters but sometimes more altitude is permitted. Therefore, we did not follow these restrictions in this exploratory study.}. To find the optimal trajectory of \acrshort{fbss}, we first must find the optimal positions of \acrshort{fbss} in different snapshots.
We consider orthogonal frequency reuse to avoid interference between \acrshort{fbss} in the network, and we consider the constraint on the number of communication channels in the intracellular network. We change the proposed mathematical model in \cite{Rahimi2021} to find the optimal position of \acrshort{fbss} in each snapshot. We consider non-line-of-sight (NLoS) path loss and aim to cover all users in each snapshot. To find the optimal trajectory of \acrshort{fbss}, we propose a mathematical model based on a transportation problem to minimize the total distance tracked by \acrshort{fbss}. We solve the proposed mathematical model for transiting \acrshort{fbss} between two snapshots in each step. We find the shortest path of each FBS while taking obstacles and potential collisions into account. We also consider that users may be situated at different altitudes, following a Poisson point process (PPP) distribution, with their mobility following a random waypoint. The \acrshort{fbss} battery and flight limitations are also considered, and we introduce an FBS set management (FSM) approach to avoid losing energy in idle hover mode and tackle the energy problem.
This article provides a mathematical model and algorithms for FBSs positioning and trajectory by considering real world challenges. Following items are the contributions of this work in comparison with the studies in the literature. 
\begin{itemize}
    \item We propose a mathematical model to solve the $3$D multi-\acrshort{fbs} positioning and trajectory problems, where NLoS links and obstacles are considered. The model has a global solution (see Table \ref{t:tablelit}).
    \begin{itemize}
        \item Using a Lemma, it is proved that the solution of the proposed algorithm is global and definite. 
    \end{itemize}
     \item We consider different users' altitudes (for example, in urban areas at ground level and on the upper floors of high-rise buildings) in the positioning and trajectory problems.
    \item To make the problem solvable, the time is considered discrete. An efficient time period for each snapshot is calculated which is not studied in FBS trajectory field (to the best of our knowledge). 
   \item We propose an FSM technique to avoid losing energy in idle modes. FBSs can fly to a base for recharge and come back to perform their tasks. 
    \item We consider power consumption for the hovering and trajectory of FBSs in the proposed model. Alongside the FSM, this helps to prolong network lifetime. 
    \item We find the shortest path for each \acrshort{fbs} avoiding obstacles and collision with each other.
\end{itemize}

The rest of the paper is as follows. The literature is reviewed in Section \ref{literature}. In Section \ref{Sys}, the system model and the channel model, is discussed. Section \ref{probformul} includes the problem formulation and proposed positioning and trajectory mathematical models. The proposed algorithm for solving the whole trajectory problem is presented at the end of this section. In Section \ref{numResults}, the test system's parameters and the numerical results are discussed. Finally, the conclusion is in Section \ref{conclusion}.

\section{Related Works} \label{literature}
Due to the high mobility, maneuverability, adaptive altitude, and low cost of FBSs, they have vital applications in wireless networks. One of the main advantages of using \acrshort{fbss} is that they do not need any pre-established infrastructure and can be deployed anywhere. They can also change their positions on-demand to increase coverage and QoS for users. However, the use of \acrshort{fbss} also has challenges, such as determining the optimal positioning, trajectory design, and number of UAVs \cite{fotouhi2019survey}, \cite{mozaffari2019tutorial}.

Crucial element of FBS-mounted wireless networks is trajectory design. An \acrshort{fbs}'s trajectory involves passing through several points obtained from the deployment problem. The literature on $2$D and $3$D positioning problems was discussed in \cite{Rahimi2021}, \cite{Sobouti2020}. 

Designing an optimal trajectory is challenging because there are an unlimited number of optimization variables (e.g., \acrshort{uavs}' positions)\footnote{Instead of the terms UxNB and FBS, the term most commonly used in the literature is UAV. In this section, we use these terms interchangeably.} \cite{zeng2016wireless}. 
The authors of \cite{qian2022path} proposed a single UAV to function as a mobile server, offloading computation tasks for a group of mobile users on the ground who move according to a random waypoint model. The solution they proposed aimed to aims to maximize average throughput while keeping energy consumption and customer fairness in mind, and their proposed time-saving Monte Carlo tree search (MCTS) algorithm was able to help them achieve that goal.
In \cite{Zhou2021}, a \acrshort{uav} path planning was proposed on the basis of the bat algorithm. The paper's primary goal was to enable UAVs to find a safer and shorter path without crashing through the start and end points in a war operation environment. The authors of \cite{Pan2021} proposed a deep learning algorithm trained by a genetic algorithm (GA). The \acrshort{ga} collected states and paths from different scenarios and then used them to train a deep neural network so that when faced with familiar scenarios, it could quickly provide an optimized path. 

In \cite{Xia2021}, a multi-\acrshort{uav} trajectory optimization model was proposed. The model was based on a single time interval and used time segmentation instead of traditional station segmentation, which simplified the calculation of cost functions. 
The authors of \cite{tang2021joint} aimed to use surveillance \acrshort{uavs} to compensate for the shortcomings of fixed surveillance systems, like CCTV.
In \cite{Zhou2020}, a secure cognitive \acrshort{uav} communication network was studied using the trajectory and high flexibility of a UAV and the possibility of creating direct vision links. 
The authors of \cite{Ji2020} investigated the problem of safe transmission in a cache-enabled \acrshort{uav} relay network with device-to-device communications, assuming the presence of a listener. 
In \cite{Ji2020cache}, the authors aimed to maximize users' minimum power by optimizing the \acrshort{uav} trajectory and transmit power. The cache, trajectory, and transmit power optimization variables were alternatively optimized in three different blocks.
The authors of \cite{samir2019trajectory} investigated the use of \acrshort{uav}s to provide a joint service to many automobiles on a highway with no infrastructure. They limited the number of \acrshort{uav}s installed, with limits on the required QoS, drawing on the amount of data and the influence of vehicle movement. They did this by using the optimal travel path of \acrshort{uav}s and assigning spectrum resources for a period of time.

In \cite{Hua2020} the goal was to maximize system power by jointly optimizing a UAV's $3$D trajectory, communication scheduling, and transmit power. In doing so, the authors first considered a specific case where the path between a UAV base station (UAV-BS) and a UAV access point (UAV-AP) is predetermined. Subsequently, they proposed an efficient iterative algorithm to optimize the $3$D UAV trajectory, and this algorithm alternately optimized the sub-problems on the basis of a sequential convex approximation technique. 
The authors of \cite{wang2020joint} proposed an architecture of relay \acrshort{uav}s to load data from smartphones onto satellites in low-Earth orbit. In so doing, they improved smartphone connection time, power management, and \acrshort{uav} trajectory to increase network capacity. Their approach involved using non-linear integer programming (NLIP) to simulate the issue.
In \cite{Feng2020}, a UAV trajectory optimization problem was formulated as a non-convex problem, which took UAV altitudes and wireless coverage performance into consideration. The authors proposed an iterative algorithm with low complexity to tackle this problem, breaking down the main problem into four sub-problems and optimizing the variables. First, a convex minimization algorithm was used to find the optimal global $2$D position of the \acrshort{uav}. Next, the optimal altitude of the \acrshort{uavs} was obtained. Then, a multi-objective evolutionary algorithm based on a decomposition algorithm was proposed to control the phase of the antenna elements and achieve the desired performance. Finally, with the variables solved, the main problem was reformulated as a single-variable optimization problem, in which charging time was the optimization variable, and the problem was solved using convex optimization techniques.
In \cite{zhang2021trajectory}, the focus was on UAV-enabled emergency networks, where \acrshort{uav}s functioned as FBSs to collect data from terrestrial users in disaster-affected areas. The energy available of the users' devices are insufficient because of the failure of the ground power supply induced by disasters. Terrestrial impediments were also shown to impact \acrshort{uav} flying due to post-disaster environmental conditions. To address the issue, the authors formulated a UAV trajectory optimization problem with user-device energy constraints and the location of terrestrial obstacles to maximize the uplink efficiency of \acrshort{uav} networks during the flight duration. They transformed the problem into a constrained Markov decision-making process (CMDP) using the \acrshort{uav} as an agent because of the dynamic user-device energy constraint. They introduced a \acrshort{uav} path-planning algorithm based on a safe deep Q-network (safeDQN) to address the CMDP problem, in which the \acrshort{uav} learned to choose optimal actions on the basis of rational policies.

The authors of \cite{ding20203d} addressed the topic of $3$D \acrshort{uav} trajectory and spectrum allocation, considering \acrshort{uav} power consumption and fairness to terrestrial users. To do this, they first defined \acrshort{uav} power consumption as a function of $3$D mobility. Then, given the restricted energy, the fair throughput was maximized. They suggested a novel Deep Reinforcement Learning-based algorithm (DRL). The proposed method allows the \acrshort{uav} to control its speed and direction to save energy and arrive at the target destination while still having enough energy and allocating the spectrum band to reach the fairness.
In \cite{You2019}, a \acrshort{uav} was used in a wireless sensor network to collect data from multiple sensor nodes (SNs). The goal a UAV was used from SNs along with the $3$D UAV trajectory. In the proposed system, the disconnection effect was also considered, and the data rate of data collection was modeled according to the communication channel. To solve the problem, the $3$D motion of the path was optimized repeatedly, once horizontally and then vertically. 
The authors of \cite{zeng2018trajectory} studied a UAV that sent a shared file to a set of ground terminals. Their goal was to optimize the path to minimize the \acrshort{uav}'s mission time. The altitude was assumed to be a constant value equal to the lowest possible altitude for a safe flight. The resulting path consisted of some waypoints and extended the traveling salesman problem, except there was no need to return to the starting point. The problem was solved so that the paths were designed to meet the minimum connection time limit during which the horizontal distance between the \acrshort{uav} and the terminal was less than a specific  value. For the given path points, the optimal speed of the \acrshort{uav} was obtained by solving a linear programming problem. In \cite{Zhan2018}, a path for a \acrshort{uav} was designed to collect data from the possible sensors. The objective was to maximize the data collected using the traveling salesman problem and convex optimization. 
The authors of \cite{wang2021robust} examined two types of \acrshort{uav}s in a \acrshort{uav}-assisted secure network. One \acrshort{uav} flew about transmitting secret data to a mobile user, while the second \acrshort{uav}, which was there to help, made bogus noises to distract the attackers. Given the mobility of \acrshort{uav}s and users, the authors aimed to increase the worst-case secrecy rate of mobile users. The challenge was handled by optimizing the $3$D trajectory of \acrshort{uav}s while taking as constraints time allocation, maximum speed, collision avoidance, placement error, and energy harvesting.

\begin{table*}[t]
\large
\caption{A comparison with existing literature}
\label{t:tablelit}
\resizebox{\textwidth}{!}{%
\begin{tabular}{|l|c|c|c|c|c|c|c|c|c|c|c|}
\hline
 &
  \cite{Wu2018} &
  \cite{li2019joint} &
  \cite{gong2018flight} &
  \cite{li2019board} &
  \cite{wang2019energy} &
  \cite{huang2019deep} &
  \cite{Hua2020} &
  \cite{zhan2019aerial} &
  \cite{samir2020age} &
  \cite{9701330} &
  Our work \\ \hline
Trajectory design      & * & * &   &   & * & * & * & * & * & * & * \\ \hline
3D trajectory          &   & * &   &   &   &   & * &   &   & * & * \\ \hline
Uplink                 &   &   & * & * & * &   & * & * & * & * &   \\ \hline
Downlink               & * &   &   &   &   &   & * &   &   &   & * \\ \hline
Sum-rate maximization  & * &   &   & * &   &   & * &   & * & * & * \\ \hline
Energy optimization    &   &   &   &   & * &   & * & * &   &   & * \\ \hline
Obstacle consideration &   &   &   &   &   & * &   &   &   &   & * \\ \hline
Time minimization      &   & * & * & * &   &   &   &   &   & * &   \\ \hline
Dynamic environment    &   &   &   &   & * &   &   & * & * & * & * \\ \hline
Mathematical solution  & * & * & * &   & * &   & * &   &   &   & * \\ \hline
\end{tabular}%
}
\end{table*}

The authors of \cite{Wu2018} jointly optimized user scheduling and \acrshort{uav} trajectories to maximize average data rates among ground users. They envisioned a wireless communication system where the UAVs served several ground users. The UAVs operated periodically, and each \acrshort{uav} had to return to the starting point at the end of each time interval. The UAV trajectories were also designed such that they respected speed limits and avoided collisions.
In \cite{li2019joint} a UAV-aided data collection
is proposed to gather data from a number of ground users.
The objective of the the paper is to optimize the UAV’s trajectory, altitude, velocity, and data links with ground users to minimize the total mission time. The paper targets emergency applications, where the mission completion time should be the main concern.
The authors of \cite{gong2018flight} considered a scenario where an UAV collects data from a set of sensors on a straight line. They considered UAV can either cruise or hover while communicating with the sensors. The objective of the paper is to minimize the UAV’s total flight time from a starting point to a destination while
allowing each sensor to successfully upload a certain amount of
data using a given amount of energy. 
In \cite{li2019board} an on-board deep Q-network is proposed to
minimize the overall data packet loss of the sensing devices. The authors have done it by optimally deciding the device to be charged and interrogated for data collection, and the instantaneous patrolling velocity of the UAV.
The authors of \cite{wang2019energy} proposed a novel UAV-assisted IoT network, in which a low-altitude UAV platform is employed as both a mobile data collector and an aerial anchor node to assist terrestrial BSs in data collection and device positioning. Theirs aim is to minimize the maximum energy consumption of all devices.
In \cite{huang2019deep} the state-of-the art deep reinforcement learning is merged with the UAV navigation through massive multiple-input-multiple-output (MIMO) technique to design a deep Q-network for optimizing the UAV trajectory by selecting the optimal policy. 
The authors of \cite{zhan2019aerial} considered a scenario where multiple UAVs collect data from a group of sensor nodes on the ground. They study tradeoff between the aerial cost, which is defined by the propulsion energy consumption and operation costs of
all UAVs, and the ground cost, which is defined as the energy
consumption of all sensor nodes. The aim is to minimize the weighted sum of the above two costs, by optimizing the UAV trajectory jointly with wake-up time allocation, as well as the transmit power of all sensor nodes. 
In \cite{samir2020age} an UAV-assisted single-hop vehicular network is considered, wherein sensors on vehicles generate time sensitive data streams, and UAVs are used to collect and process this data while maintaining a minimum age of information.
The authors of \cite{9701330} focused on the problem of deploying
UAV-BSs to provide satisfactory wireless communication services,
with the aim to maximize the total number of covered user equipment subject to user data-rate requirements and UAV-BSs’ capacity
limit.

The authors of \cite{hu2021distributed} investigated how to construct a trajectory for a group of energy-constrained \acrshort{uav}s working in dynamic wireless network situations. In their model, a group of drone base stations (DBSs) was dispatched to jointly service clusters of ground users with dynamic and unexpected uplink access requests. In this scenario, the DBSs had to maneuver together to maximize coverage for the dynamic requests of ground users. This optimization approach for trajectory design aimed to develop optimal trajectories that increased the fraction of customers served by all FBSs. A value decomposition-based reinforcement learning (VD-RL) method with a meta-training mechanism was proposed to obtain an optimal solution for this non-convex optimization problem in unpredictable situations.
In \cite{Tang2019} the objective was to minimize the average power through both the design of transmission power and the travel path for an activated network of \acrshort{uavs}. The authors proposed a new alternative optimization method by combining power and trajectory in an intermediate variable and then updating the power and the newly introduced variable. This new variable simplified the analysis of the main problem by turning it into two convex sub-problems, namely an operational power maximization sub-problem and a feasibility sub-problem.
In \cite{9701330}, the authors developed a UAV-assisted IoT system that maximized the quantity of data gathered from IoT devices while depending on the UAVs' shortest flight paths. After that, a method based on deep reinforcement learning was developed to determine the best trajectory and throughput within a given coverage region. Following training, the UAV was able to gather all the data from user nodes independently, improving the overall sum rate while utilizing fewer resources. 
Based on a connected graph, \acrshort{uav} routes were designated to serve IoT devices in \cite{kouroshnezhad2020energy}. Their proposed method, known as semi-dynamic mobile anchor guidance (SEDMAG), used a weighted search algorithm to determine the shortest-path energy for conservative planning to meet the nodes dynamically.

In this article, several important issues have been considered, which, to the best of our knowledge, have not been addressed independently or jointly in other related works. The most important contribution in the proposed algorithm is to provide an energy management method for FBSs, called FSM, which makes the operation time of FBSs longer and makes the problem closer to reality. In the FSM method, in each snapshot, idle FBSs return to the base and recharged so that they can be used again if needed in the future. Moreover, a method that finds the shortest possible 3D path for each FBS between any two snapshots is presented, avoiding obstacles and collisions between FBSs. In appendix, it is proved that the path is the shortest path and the optimal solution to the problem. In addition to points mentioned, we have provided a solution to find the right duration for each snapshot, which is the right time to recheck the status of network users according to the conditions of the problem. Furthermore, to find the optimal 3D positions of FBSs in each snapshot, we have presented a mathematical model with an optimal solution in the problem space, considering the path loss and the altitude of the users.

In this paper, we have tried to be the closest to the real world conditions compared to the related works. We propose an exact solution in the problem space in both positioning and trajectory phases, while taking constraints related to power consumption, data rate, interference, and FBS collision avoidance into account. Moreover, we have used mathematical model to have definite and global solution compared to heuristics or learning algorithms, which are time-consuming and do not produce an exact solution.

\section{System Model}\label{Sys}
\begin{figure}[t]
        \centering
        \includegraphics[width=0.9\columnwidth]{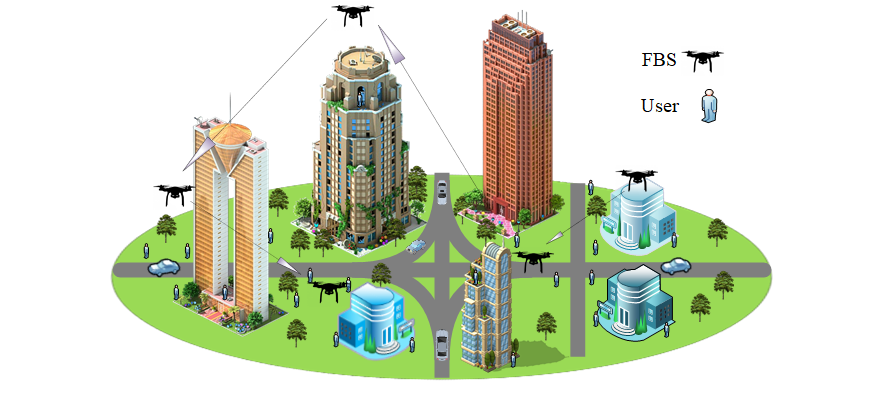}
        \caption{A possible scenario for the proposed system.}%
        {{\small }}    
        \label{system_fig}
\end{figure}

In this paper, we consider a wireless cellular network in an urban environment. Users are present and gathered in different parts of the area at different times, so it is necessary to dynamically change the location of the base stations to provide the best possible services to users. Since deploying a terrestrial base station in short-term scenarios is not economically viable, we aim to offer coverage and service to all cellular users by means of FBSs.  It is assumed that users in this system have different data rate needs, which we have assumed to be random on the basis of uniform distribution. Also, communication links between users and \acrshort{fbss} are assumed to be both \acrshort{los} and \acrshort{nlos}.
A possible assumed system is shown in Figure \ref{system_fig}.
\par Our main goal in this paper is to solve a $3$D trajectory planning problem for FBSs that would provide coverage and service to users with the fewest possible FBSs. To do this, we have divided the problem into several snapshots. In each snapshot, we find the most suitable $3$D positions and the minimum possible number of \acrshort{fbss}. Then, we solve the trajectory problem to minimize the energy consumption of \acrshort{fbss}.
\par We solve the $3$D positioning problem of \acrshort{fbss} in each snapshot based on the proposed model in \cite{Rahimi2021} considering \acrshort{nlos} links and assuming $100$ percent users coverage. In doing so, we find the minimum number of FBSs required on the basis of the proposed bisection algorithm. The altitude of \acrshort{fbss} can be assumed between the two values of $\rm{H}_{min}$ and $\rm{H}_{max}$ based on the characteristics of the \acrshort{fbss} used. In addition, different altitudes have been assumed for users in this problem, which represents the presence of users in high-rise buildings. In order to prevent inter-cell interference, we have considered the limitation of the number of internal channels in the cell.
\par To solve the trajectory problem, we find the shortest path between each point of origin and the destination between two consecutive snapshots while taking obstacles and the potential for collisions with FBSs into consideration.
To minimize the \acrshort{fbss}' energy consumption, we propose a linear mathematical model based on the transportation problem between two snapshots, whereby the optimal trajectory of each FBS from origin to destination is obtained. We also assume that users move based on a random waypoint. In the following section, this will be discussed in more detail. First, however, we should introduce the communication channel between the \acrshort{fbs} and the user.

The deployment and service of \acrshort{fbss} are directly related to the \acrfull{a2g} communication links. Various models for the \acrshort{a2g} channel have been introduced in the literature. In this paper, we use the A2G model presented in \cite{bor2016efficient}. The channel model generally consists of two parts: \acrshort{los} and \acrshort{nlos} links. The possibility of an \acrshort{los} link between the \acrshort{fbs} and the ground user is determined by several parameters, including building density, \acrshort{fbs} location, and the elevation angle between the \acrshort{fbs} and the ground user. In an \acrshort{a2g} channel, the probability of an \acrshort{los} link is calculated as follows:
\begin{equation}\label{P_LoS}
P_{\mathsf{LoS}}=\frac{1}{1+a \rm exp \textit{$(-b(\theta_{u} - a))$}},
\end{equation}
where $a$ and $b$ are environmental constants, and $\theta_{u}$ is the elevation angle between user $u$ and the \acrshort{fbs}, which depends on their altitudes. It is calculated as $\theta_{u} = (\frac{-180}{\pi}tan^{-1} \frac{h_{\mathsf{FBS}} - h_{u}}{d_{u}})$, where $h_{\mathsf{FBS}}$ and $h_{u}$ are the altitudes of \acrshort{fbs} and user $u$, respectively. Also, $d_{u}$ is the distance between the \acrshort{fbs} and user $u$. It is calculated as $d_{u} = \sqrt{(x_{u} - x_{\mathsf{FBS}})^2 + (y_{u} - y_{\mathsf{FBS}})^2 + (h_{u} - h_{\mathsf{FBS}})^2 } $. It can be seen from \eqref{P_LoS} that the probability of an LoS link increases in accordance with the increasing elevation angle between the \acrshort{fbs} and the user. The probability of having an \acrshort{nlos} communication link can be calculated as follows:
\begin{equation}\label{P_NLoS}
P_{\mathsf{NLoS}}=1 - P_{\mathsf{LoS}}.
\end{equation}
Hence, the mean path loss (in dB) for \acrshort{los} and \acrshort{nlos} communication links can be calculated accordingly \cite{kalantari2020wireless}:
\begin{equation}\label{L_LoS}
L_{\mathsf{LoS}}= 20 \log(\frac{4\pi f_{c} d_{u}}{C}) + \delta_{\mathsf{LoS}} ,
\end{equation}
\begin{equation}\label{L_NLoS}
L_{\mathsf{NLoS}}= 20 \log(\frac{4\pi f_{c} d_{u}}{C}) + \delta_{\mathsf{NLoS}} ,
\end{equation}
where $\delta_{\mathsf{LoS}}$ and $\delta_{\mathsf{NLoS}}$ are mean losses in \acrshort{los} and \acrshort{nlos} communication links, respectively. Also, $C = 3 \times 10^8 $ is the light speed, and $f_{c}$ is the carrier frequency. Therefore, the probabilistic long-term mean path loss is obtained as
\begin{equation}\label{L_du}
L(d_{u})= L_{\mathsf{LoS}} \times P_{\mathsf{LoS}} +  L_{\mathsf{NLoS}} \times P_{\mathsf{NLoS}}.
\end{equation}

\section{Problem Formulation} \label{probformul}
In aiming to solve the problem of $3$D trajectory planning for FBSs in a cellular network, we must find an unlimited number of continuous points for the FBSs' positions, which is an NP-hard problem \cite{Rahimi2021}. To do this, we divide the problem into several snapshots and find the optimal path of \acrshort{fbss} between every two snapshots, taking FBS energy consumption into consideration. We divide the problem into two phases. First, we find the minimum number of FBSs required and their optimal positions in each snapshot to cover and serve users. Then, we find the optimal path of each \acrshort{fbs} from the origin to the destination. 
\subsection{Positioning phase}
\par In the first phase, the objective function of the positioning problem is to minimize the number of \acrshort{fbss} required and and to find the optimal positions of \acrshort{fbss} to cover users. To do this, we reformulate the method proposed for FBS positioning in \cite{Rahimi2021}. We consider \acrshort{nlos} links in this problem and aim to cover all users. We also consider \acrshort{fbs} capacity to avoid inter-cell interference between covered users.

\begin{table}[t]
\caption{Decision variables in the positioning phase.}
\resizebox{\columnwidth}{!}{%
\centering
\begin{tabular}{c|l}
Decision variable                    & \multicolumn{1}{c}{ Description}                                       \\ \hline
$x_{ij}$                    & $1$, if user $j$ is served by candidate point $i$,\\
&and $0$, otherwise.                             \\
$m_i$                & $1$, if candidate point $i$ is selected for \acrshort{fbs}\\
&deploying, and $0$, otherwise. \\
$h_i$                   &The altitude of the \acrshort{fbs} which is deployed at the\\
&candidate point $i$.\\
$k_{ij}$               &The path loss between user $j$ and candidate\\
&point $i$, if user $j$ is served by candidate point\\ 
&$i$, and $0$, otherwise.\\
$t_{ij}$              &Auxiliary decision variable.
\end{tabular}
}
\label{decision_variable_tbl}
\end{table}

\par The decision variables considered in this formulation are presented in Table \ref{decision_variable_tbl}. $x_{ij}$ is the variable that represents whether or not user $j$ is served by an FBS at candidate point $i$. $m_{i}$ shows whether candidate point $i$ is selected or not. $h_{i}$ is the variable of the \acrshort{fbs}'s altitude. The path loss between user $j$ and the \acrshort{fbs} at candidate point $i$ is decided by variable $k_{ij}$. The $t_{ij}$ is an auxiliary decision variable. Also, the parameters used in this formulation are represented in Table \ref{Parameters_tbl}.


\begin{subequations}
\begin{alignat}{1}
& {\rm min} \sum_{i \in \mathcal{I}}\sum_{j \in \mathcal{J}} k_{ij} \label{Lobj2}\\
&{\rm s.t} \cr
&\quad \sum_{i \in \mathcal{I}} x_{ij} \leq 1, \quad \forall j \in \mathcal{J}, \label{LC1} \\
&\quad \sum_{j \in \mathcal{J}} x_{ij} \leq \psi_{\mathsf{FBS}}   , \quad \forall i \in \mathcal{I}, \label{LC1_1} \\
& \quad x_{ij} \leq m_i, \quad \forall i \in \mathcal{I} ,j \in \mathcal{J}, \label{LC2}  \\
&\quad \sum_{i \in I\mathcal{I}} \sum_{j \in \mathcal{J}} x_{ij} =  U, \label{LC3}\\
&\quad \sum_{j \in \mathcal{J}} D_j \times x_{ij} \leq \beta \times m_i, \quad \forall i \in \mathcal{I}, \label{LC4} \\
&\quad \sum_{i \in \mathcal{I}} m_i = P, \label{LC5}\\
&\quad h_i \leq H_{{\rm max}} \times m_i, \quad \forall i \in \mathcal{I}, \label{LC6}\\
&\quad h_i \geq H_{{\rm min}} \times m_i, \quad \forall i \in \mathcal{I}, \label{LC7}\\
&\quad \cot(\theta) \times x_{ij} \leq \frac{h_i}{d_{ij}},\quad \forall i \in \mathcal{I} ,j \in \mathcal{J}, \label{LC8}  \\
&\quad  k_{ij} \geq [P_{\mathsf{LoS}} \times (4{\pi} \frac{f_c}{C})^2 d_{ij}^2   -h_0^2] x_{ij}\cr
& \quad \quad +(4{\pi} \frac{f_c}{C})^2 \times 2 \times h_0 \times t_{ij} + x_{ij} \times 10^{\delta_{\mathsf{LoS}}} \cr
& \quad \quad + [P_{\mathsf{NLoS}} \times (4{\pi} \frac{f_c}{C})^2 d_{ij}^2 -h_0^2] x_{ij} \cr
& \quad \quad  +(4{\pi} \frac{f_c}{C})^2 \times 2 \times h_0 \times t_{ij}+ x_{ij} \times 10^{\delta_{\mathsf{NLoS}}}, \cr
&\quad \quad  \quad  \quad  \quad  \quad  \quad  \quad  \quad  \quad  \quad  \quad  \quad  \quad   \forall i \in \mathcal{I} ,j \in \mathcal{J},  \label{LC11}\\
&\quad k_{ij} \leq M \times x_{ij}, \quad \forall i \in \mathcal{I} ,j \in \mathcal{J}, \label{LC10}  \\
&\quad  t_{ij} \leq h_i, \quad \forall i \in \mathcal{I}, j \in \mathcal{J}, \label{LC12}\\
&\quad t_{ij} \leq H_{{\rm max}} \times x_{ij}, \quad \forall i \in \mathcal{I}, j \in \mathcal{J}, \label{LC13}\\
&\quad  t_{ij} \geq h_i - (1-x_{ij})H_{{\rm max}},\quad \forall i \in \mathcal{I}, j \in \mathcal{J}. \label{LC14}
\end{alignat}
\end{subequations}

In the proposed model, the objective function (\ref{Lobj2}) is defined to minimize the sum of path losses. Constraint (\ref{LC1}) stipulates that each user must be served by only one \acrshort{fbs}. Constraint (\ref{LC1_1}) states that each \acrshort{fbs} can serve a limited number of users based on its number of channels. Constraint (\ref{LC2}) shows that user $j$ can only be served by the \acrshort{fbs} deployed at candidate point $i$. Constraint (\ref{LC3}) stipulates that \acrshort{fbss} must cover all users. Constraint (\ref{LC4}) allows each \acrshort{fbs} to serve its maximum data rate based on its backhaul. Constraint (\ref{LC5}) states that the model must select only $P$ points from the given candidate points. Constraints (\ref{LC6}) and (\ref{LC7}) stipulate that if candidate point $i$ is selected as the position of an \acrshort{fbs}, the \acrshort{fbs} must fly within the permissible range. The \acrshort{fbs} altitude will be set to zero if the model does not select the candidate point $i$. Constraint (\ref{LC8}) prevents the assignment of users who are not in the \acrshort{fbs}'s coverage range. Constraints (\ref{LC11}) and (\ref{LC10}) are the first-order Taylor expansion of equation \ref{L_NLoS}, proven in Lemma \ref{lem}.
In constraints (\ref{LC12})--(\ref{LC14}) the decision variable $t_{ij} = x_{ij} \times h_{i}$ is used to reduce the nonlinear part to the multiplication of $x_{ij}$ and $h_{i}$. $t_{ij}$ must be zero if $x_{ij}$ or $h_{i}$ are equal to zero. Constraints (\ref{LC12}) and (\ref{LC13}) state this requirement. Also, $t_{ij}$ must be equal to $h_{i}$ when $x_{ij}$ becomes $1$. Constraints (\ref{LC12}) and (\ref{LC14}) satisfy this.

\begin{lemma}\label{lem}
 Consider $L(d_{ij})= L_{\mathsf{LoS}} \times P_{\mathsf{LoS}} +  L_{\mathsf{NLoS}} \times P_{\mathsf{NLoS}}$ as the path-loss function. If $L(d_{ij}) \geq {\mathsf{PL_{max}}}$, then $x_{ij}$ must be equal to $0$. The statement can be rewritten as follows:
 
 \begin{equation}\label{ReCondition}
 x_{ij}= \left\{
\begin{array}{ll}
0, & \text{if } L(d_{ij}) \geq \mathsf{PL_{{max}}},\\
0 \quad  {\rm or} \quad 1, & \text{otherwise}.\\

\end{array} \right.
 \end{equation}

\end{lemma}
\begin{proof}
We obtain a linear conditional statement in terms of $h_i$ by replacing $L(d_{ij})$ in \eqref{ReCondition} with its linear approximation achieved  from Taylor expansion around some $h_0$:
  \begin{multline*}
 L(d_{ij}) = L(h_i - h_0 + h_0) \approx L(h_0) + L^{'} ( h_0) (h_i -h_0) \\
 = [P_{\mathsf{LoS}} \times (4{\pi} \frac{f_c}{C})^2 d_{ij}^2   +h_0^2] x_{ij}\quad \quad \quad \quad \quad \quad \quad \quad  \\
 +(4{\pi} \frac{f_c}{C})^2 \times 2 \times h_0 \times (h_i - h_0) + x_{ij} \times 10^{\delta_{\mathsf{LoS}}} \\
 + [P_{\mathsf{NLoS}} \times (4{\pi} \frac{f_c}{C})^2 d_{ij}^2 +h_0^2] x_{ij} \quad \quad \quad\quad\quad\quad\\
  +(4{\pi} \frac{f_c}{C})^2 \times 2 \times h_0 \times (h_i - h_0)+ x_{ij} \times 10^{\delta_{\mathsf{NLoS}}}.   
 \end{multline*}
 Now, we have
  \begin{equation}\label{ReCondition1}
  x_{ij}= \left\{
\begin{array}{ll}
0, & \text{if } 
[P_{\mathsf{LoS}} \times (4{\pi} \frac{f_c}{C})^2 d_{ij}^2   +h_0^2] x_{ij}
\\
 &+(4{\pi} \frac{f_c}{C})^2 \times 2 \times h_0 \times (h_i - h_0) 
 \\&+ x_{ij} \times 10^{\delta_{\mathsf{LoS}}} 
 \\
& + [P_{\mathsf{NLoS}} \times (4{\pi} \frac{f_c}{C})^2 d_{ij}^2 +h_0^2] x_{ij} 
 \\
 & +(4{\pi} \frac{f_c}{C})^2 \times 2 \times h_0 \times (h_i - h_0)\\
 &+ x_{ij} \times 10^{\delta_{\mathsf{NLoS}}}. \\
 &\geq  \mathsf{PL_{{max}}},\\
0 \quad  {\rm or}  \quad 1, & \text{otherwise}.\\
\end{array}\right.
 \end{equation}
 Assume that $A = (4{\pi} \frac{f_c}{C})^2$. By simplifying the conditional expression, we have
   \begin{equation*}
     x_{ij}= \left\{
\begin{array}{ll}
0, & \text{if } 
h_i \geq \frac{\mathsf{PL_{{max}}} -(A\times (d_{ij}^2 -h_0^2) \times(P_{\mathsf{LoS}} + P_{\mathsf{NLoS}}) }{2Ah_0},\\
0 \quad  {\rm or}  \quad 1, & \text{otherwise.}\\
\end{array} \right. 
   \end{equation*}
   By defining $a_{ij}=\frac{\mathsf{PL_{{max}}} -(A\times (d_{ij}^2 -h_0^2) \times(P_{\mathsf{LoS}} + P_{\mathsf{NLoS}}) }{2Ah_0}$, the conditional expression will be simplified as follows:
   
   \begin{equation}\label{ReCondition2}
        x_{ij}= \left\{
\begin{array}{ll}
0, & \text{if } 
h_i \geq a_{ij},\\
0 \quad  {\rm or}  \quad 1, & \text{otherwise.}\\
\end{array} \right.
 \end{equation}
 To form \eqref{ReCondition2} as a valid constraint in mathematical programming, the following expression can be given:
 \begin{equation}\label{ReCC9}
 x_{ij} \leq \frac{M - h_i}{M-a_{ij} + \frac{1}{2}}, \quad \forall i \in \mathcal{I} ,j \in \mathcal{J},
 \end{equation}
where $M$ is a large number. 

\end{proof}


\begin{tiny}
\begin{table}[t]
\caption{Parameters used in the positioning phase.}
\resizebox{\columnwidth}{!}{%
\centering
\begin{tabular}{c|l}
Parameters                      & \multicolumn{1}{c}{Description}                                       \\ \hline
$f_c$                  &Carrier frequency\\
$C$                    &Speed of light\\
$\mathcal{I}$                    & Set of candidate points \\
$\mathcal{J}$                & Set of users \\
$\beta$                     & \acrshort{fbs} backhaul data rate \\
$\theta$                &\acrshort{fbs} elevation angle\\
$P$                     & Number of \acrshort{fbss} to be deployed \\
$U$             & Number of users                     \\
$H_{{\rm min}}$                &Minimum allowed altitude\\
$H_{{\rm max}}$                &Maximum allowed altitude\\
{$\mathsf{PL_{max}}$ }  & Maximum allowed path loss in the network\\
{$\psi_{\mathsf{FBS}}$}            & Number of inter-cell channels \\
$D_j$                     & Mean data rate required by user $j$\\   
$d_{ij}$                     & Distance between user $j$ and candidate point $i$                        
\end{tabular}
}
\label{Parameters_tbl}
\end{table}
\end{tiny}


\subsection{Trajectory phase}
In the trajectory phase, our objective is to find the best path for each \acrshort{fbs} between two snapshots while taking obstacles and the potential for collisions into account. To do so, we must first consider the interval between every two snapshots. Since the velocity of \acrshort{fbss} is considered constant, the time between every two snapshots can be also constant. According to the characteristics of the \acrshort{fbs}, a constant speed of $15$ meters per second (m/s) has been considered for the \acrshort{fbs}. This speed is also acceptable considering the average speed of users in the urban cellular network, including fixed users, pedestrian users (with an average speed of $1$ m/s), and vehicular users (with an average speed of $10$ m/s).
According to the (\ref{delta_t}), the $\Delta t$ can be obtained on the basis of the average speed of users ($\Gamma$) and the minimum coverage range of an \acrshort{fbs}. This means, on average,  $\Delta t$ seconds for a user to move out of the minimum coverage range of an FBS:
\begin{equation}
    \Delta t = \frac{R_{\rm min}}{\Gamma}. 
\label{delta_t}
\end{equation}

The average speed of users ($\Gamma$) can also be obtained from \eqref{max_distance}. In this regard, it is assumed that $\alpha\%$ of users are stationary, $\beta\%$ are pedestrians, and $\gamma\%$ are moving in cars:
\begin{equation}
    \Gamma = \frac{\alpha \times V_{\alpha} + \beta \times V_{\beta} + \gamma \times V_{\gamma}}{100} ,
\label{max_distance}
\end{equation}
where $V_{\alpha}$ is equal to $0$.

After finding the hovering positions where \acrshort{fbss} hover and serve users, we tackle the problem of calculating the best \acrshort{fbss} path. The purpose of this is to discover the shortest path between the origin and destination of an FBS for every two snapshots, while avoiding obstacles such as buildings.
To do this, we first create a graph containing source and destination hovering points and obstacle edge points. As obstacle edges are continuous in $3$D, we consider points on their edges with a fixed distance to simplify the problem while keeping generality. This distance should not be so little that the graph grows too vast, causing the problem to take too long to solve. The distance should also not be so great that it substantially impacts the solution, with the result deviating from optimal global solutions. On this point, we should note that small distances of discretizing edges (e.g., less than $5$ meters) do not significantly affect the solution compared with \acrshort{fbss}' altitude and problem space.
Then, we create each edge between two graph vertices with a cost equal to the Euclidean distance in three dimensions. In the created graph, the edge between two vertices is ignored if an obstacle splits them. 
Dijkstra's algorithm \cite{dijkstra1959note} should be used to identify the shortest paths between every two hovering spots after the graph and edge cost have been determined. Dijkstra's algorithm is a graph traversal method that finds the shortest path between two specified vertices in a weighted graph with no negative edges. Therefore, an \acrshort{fbs} can fly from one position to another using this graph without colliding with obstacles, as proven in Lemma \ref{lem_dji}.
\begin{lemma}\label{lem_dji}
In graph G(V,E), when going only via visited nodes, dist[V] is the shortest distance from source to V, or infinite if no such path exists. (Note that we don't assume dist[V] is the shortest distance for nodes that haven't been visited.)
\end{lemma}
\begin{proof}
The base case is when just one node is visited, namely the initial node source, in which case the hypothesis is straightforward.

Otherwise, we use the $n-1$ visited nodes hypothesis. In such an instance, we pick an edge $V$-$U$ with the least $dist[U]$ of any unvisited nodes, such that $dist[U] = dist[V] + G.E[V, U]$. Because if there was a shorter way and $W$ was the first unvisited node on that path, the original hypothesis would stipulate that $dist[W] > dist[U]$, which is a contradiction. Similarly, if there was a shorter path to $U$ that did not use any unvisited nodes, and if the last but one node on that path was $W$, then $dist[U] = dist[W] + G.E[W, U]$ would be a contradiction.

After processing $U$, $dist[W]$ will still be the shortest distance from the source to $W$ using only visited nodes, because if there were a shorter path that did not go through $U$, we would have discovered it before processing $U$. If a shorter path did go through $U$, we would have updated it when processing $U$.

Therefore, the shortest path from the source to node $V$ consists only of visited nodes once all nodes have been visited; hence $dist[V]$ is the shortest distance.
\end{proof}
We run the shortest path algorithm for all origin-destination points. So the shortest distance between each origin-destination pair will be found.

In formulating the problem to find the best path, as mentioned in Table \ref{traj_decision_variable_tbl}, $d_{ij}$ is the decision variable that must be equal to $1$ if \acrshort{fbs} $i$ moves to position $j$; otherwise, it must be equal to $0$.
The objective function is to minimize the total energy consumed by \acrshort{fbss} due to their flight (\ref{Tobj1}). $E_{ij}$ is obtained as follows:
\begin{equation}
    E_{ij} = \mathsf{dist_{ij}} \times E_{1} + E_U,
\end{equation}
where $E_{1}$ is the energy consumption of the \acrshort{fbs} for $1$ meter of flying, which is calculated as 
\begin{equation}\label{E_1_eq}
    E_{1} = \frac{\zeta \times V \times 3600}{D},
\end{equation}
where $\zeta$ is the battery capacity, $V$ is the battery's voltage, and $D$ is the total possible distance of the \acrshort{fbs} flight.
Also, $E_U = m g \Delta h$ is the potential energy consumed for \acrshort{fbs} $\Delta h$ altitude change. Therefore, $E_{ij}$ calculated as
\begin{equation}\label{E_ij_eq}
    E_{ij} = (\mathsf{dist_{ij}} \times E_{1}) + (m g \Delta h).
\end{equation}

Constraint (\ref{TC1}) states that the model decides the path of each \acrshort{fbs} placed in the origin snapshot. The constraint (\ref{TC2}) ensures  that each position in the subsequent snapshot is chosen for one \acrshort{fbs}. $\mathcal{M}$ and $\mathcal{N}$ are sets of \acrshort{fbs} positions in the origin and destination snapshots, respectively. The size of each set is twice the maximum required \acrshort{fbss} to ensure that the mathematical model has a feasible solution in every iteration.
(\ref{TC4}) is the constraint of the \acrshort{fsm}; when the number of \acrshort{fbss} required in one snapshot is less or more than another, extra \acrshort{fbss} are moved to or from the base to save their energy, recharge, or extend the network coverage. The concept of \acrshort{fsm} is similar to that of \acrfull{cso}.
Constraint (\ref{TC5}) ensures that each \acrshort{fbs} has enough power to fly to the next position. {\sffamily Energy$_{{\mathsf{th}}}$} is the minimum energy required for each \acrshort{fbs} to fly to the base. If an \acrshort{fbs} does not have enough power to continue serving users, the $d_{ij}$ variable will choose the path to the base. The \acrshort{fbs} will return to the base, and the other \acrshort{fbs} will be used to fly from the base to the destination point. We call this process \acrshort{fsm}:

\begin{subequations}
\begin{alignat}{1}
& {\rm min} \sum_{i \in \mathcal{M}}\sum_{j \in \mathcal{N}} d_{ij} \times E_{ij} \label{Tobj1}\\
&{\rm s.t} \cr
&\quad \sum_{j \in \mathcal{N}} d_{ij} = 1, \quad \forall i \in \mathcal{M}, \label{TC1} \\
&\quad \sum_{i \in \mathcal{M}} d_{ij} = 1, \quad \forall j \in \mathcal{N}, \label{TC2} \\
& \quad \sum_{i \in \mathcal{M}}\sum_{j \in \mathcal{N}} d_{ij} = \rm max(\mathcal{M},\mathcal{N}), \label{TC4}\\
& \quad d_{ij} \times E_{ij} \leq \mathsf{Energy_{th}} , \quad \forall i \in \mathcal{M} ,j \in \mathcal{N}, \label{TC5} \\
& \quad d_{ij} \times \mathsf{dist_{ij}} \leq \mathsf{Distance_{th}} , \quad \forall i \in \mathcal{M} ,j \in \mathcal{N}. \label{TC3}
\end{alignat}
\end{subequations}
Constraint (\ref{TC3}) states that no \acrshort{fbs} can fly further than its threshold distance based on its constant velocity. The {\sffamily Distance$_{{\mathsf{th}}}$} parameter is derived as follows:
\begin{equation}\label{distance_th}
    \mathsf{Distance_{th}} = V_{\mathsf{FBS}} \times \Delta t,
\end{equation}
where  $\Delta t$ is the time interval between two snapshots. The trajectory parameters are described in Table \ref{traj_Parameters_tbl}.
\begin{table}[t]
\caption{Decision variable in the trajectory phase.}
\resizebox{\columnwidth}{!}{%
\centering
\begin{tabular}{c|l}
Decision variable                    & \multicolumn{1}{c}{ Description}                                       \\ \hline
$d_{ij}$                    & 1, if \acrshort{fbs} $i$ moves to position $j$ in the next snapshot,\\
&and $0$; otherwise.                             \\
\end{tabular}
}
\label{traj_decision_variable_tbl}
\end{table}

\begin{tiny}
\begin{table}[t]
\caption{Parameters used in the trajectory phase.}
\resizebox{\columnwidth}{!}{%
\centering
\begin{tabular}{c|l}
Parameters                      & \multicolumn{1}{c}{Description}                                       \\ \hline
$E_{ij}$                  & The energy that the \acrshort{fbs} consumes to fly between \cr
&position $i$ and position $j$\\
$\mathcal{M}$                    & Set of origin \acrshort{fbs} positions \\
$\mathcal{N}$                & Set of destination \acrshort{fbs} positions \\
$\mathsf{Energy_{th}}$   & Minimum required energy for an \acrshort{fbs} to fly to the base\\
$\mathsf{dist_{ij}}$                  & The distance between position $i$ and position $j$\\
$\mathsf{Distance_{th}} $   & Maximum allowed distance that the \acrshort{fbs} can fly \cr
&between two snapshots\\
\end{tabular}
}
\label{traj_Parameters_tbl}
\end{table}
\end{tiny}

The total energy of the path for each \acrshort{fbs} ($E_{\rm path}$) is obtained as follows:
\begin{equation}
    E_{\rm path} = E_{ij} + E_{\rm hover},
\end{equation}
where $E_{\rm hover}$ is the energy consumed when the \acrshort{fbs} hovers.
In Theorem presented in appendix \ref{appendix1}, we prove that the obtained path considering obstacles is the global shortest path possible for each FBS.

\subsection{FSM}
After finding the path energy ($E_{\rm path}$) of each \acrshort{fbs} for each subsequent snapshot, we must check if the energy of each \acrshort{fbs} is enough to fly to the destination through the selected path. Suppose the remaining energy of the \acrshort{fbs} after flying to the destination is less than $\mathsf{Energy_{th}}$, or the number of required \acrshort{fbss} in the subsequent snapshot is less than the current snapshot. In that case, some \acrshort{fbss} must return to the base to recharge and wait. Moreover, if in the subsequent snapshot more \acrshort{fbss} are needed, the \acrshort{fbss} required must fly to the selected destination points through the selected paths. This whole process is called \acrshort{fsm}

As a general solution to the entire problem, we carry out the following steps. First, we divide the problem into several snapshots. In each snapshot, users may have different positions and data rate demands. In each snapshot, we solve the $3$D positioning problem to find the minimum number of required \acrshort{fbss} and their proper positions. Then, we create a graph based on origin and destination hovering points and obstacle edge points. After that, we find the shortest path between each origin-destination pair while taking obstacles into consideration. As Theorem \ref{lem_fbs_collision} proves, the probability of an \acrshort{fbs} colliding is almost zero.
Next, we solve the trajectory problem between two snapshots in a row using the proposed mathematical model and updating the \acrshort{fbss} energy parameter considering trajectory and hovering power consumption.
The proposed trajectory model decides the best path for each \acrshort{fbs} to minimize the total energy consumption. It also decides whether an \acrshort{fbs} should return to base due to a lower number of required \acrshort{fbs} or due to the need to recharge batteries (concept \acrshort{fsm}). 
Both proposed mathematical models are in linear form. The proposed positioning model is a mixed-integer linear problem (MILP), and the proposed trajectory model is a binary linear problem (BLP). Also, Dijkstra's algorithm has an exact solution. Therefore, the proposed method reaches the exact solution.
The process continues until there is no remaining snapshot. The whole process is shown in Algorithm \ref{alg1}.
\begin{algorithm}[t]
\caption{The \acrshort{fsm} algorithm}
\label{alg1}
\begin{algorithmic} 
\STATE - Solve the positioning problem in $s_{0}$ snapshot and find the number of required \acrshort{fbss} ($\mathcal{M}$)
\STATE - Calculate \textDelta $t$ using eq. \ref{delta_t}
\WHILE {there is remaining snapshot}
\STATE - Solve the positioning problem in $i$-th snapshot ($s_{i}$) and\\ \hspace*{0.5cm} find the number of required \acrshort{fbss} ($\mathcal{N}$)
\STATE - Create graph $G(V,E)$
\STATE - Find the shortest path between every origin-destination\\ \hspace*{0.5cm} pair from $\mathcal{M}$ and $\mathcal{N}$ considering collisions
\STATE - Calculate $\mathsf{Distance_{th}}$ using eq. \ref{distance_th}
\STATE - Solve the proposed mathematical model for trajectory\\ \hspace*{0.5cm} using $\mathcal{M}$ and $\mathcal{N}$
\STATE - Update \acrshort{fbss} remaining energy
\IF{an \acrshort{fbs} landed in the base}
\STATE - recharge it
\ENDIF
\STATE - $\mathcal{M} \leftarrow \mathcal{N}$
\ENDWHILE
\end{algorithmic}
\end{algorithm}

\begin{theorem}{The probability of an \acrshort{fbs} colliding is almost zero.}

\label{lem_fbs_collision}

\end{theorem}
\begin{proof}
We know that infinite lines can be drawn in $3$D. Therefore, the probability of finite lines colliding is almost zero ($1$). 
Also, since the origin and destination points of neither of the two \acrshort{fbs} paths in the proposed deployment model are equal, there are no two paths that originate from one point ($2$).
If the destination point of one path intersects with the origin point of another, since the second \acrshort{fbs} had moved before and changed its position, the first \acrshort{fbs} will not collide with the second \acrshort{fbs} at the end of its path ($3$).
In addition, if two \acrshort{fbs} paths collide and only if the intersection point is at the same distance from the origin point of each \acrshort{fbs} (assuming the speed of both \acrshort{fbss} is constant and equal), then the \acrshort{fbss} will collide with each other. The probability of this event is almost zero, and by taking turns in the movement of \acrshort{fbss}, this event can also be prevented ($4$).

By taking $1$, $2$, $3$, and $4$ into consideration, we can conclude that the probability of a collision between the two \acrshort{fbss} in the proposed model is almost zero.
\end{proof}

\section{Numerical Results}\label{numResults}
In this section, we first introduce the test system and simulation parameters. We then discuss and compare the results of scenarios for the $3$D \acrshort{fbs} trajectory problem.

\subsection{Test system}
In the simulations, we consider a centralized decision-making  system for \acrshort{fbss} positioning and trajectory. Referring to \cite{kalantari2020wireless} and \cite{zhong2020qos}, as mentioned in Table \ref{testparameters}, we consider a $5,000 \times 5,000$ meter urban area with scenarios including $80$, $200$, and $450$ users with Poisson point process (PPP) distribution. The PPP parameter $\lambda = 20,000$, and the environment parameters are as follows: $f_c = 2$ GHz, $\mathsf{PL_{max}} = 110$ dB, and ($a$, $b$, $\delta_{\mathsf{LoS}}$, $\delta_{\mathsf{NLoS}}$) $=$ ($9.61$, $0.16$, $1$, $20$), corresponding to urban environments.
Additionally, we consider that the data rate requested by users has a uniform distribution with a maximum value of $6$ Mbps ($D_{avg} = 3$ Mbps). The \acrshort{fbss}' backhaul data rate ($\beta$) is considered equal to $100$ Mbps for each. The \acrshort{fbss}' flying altitude range is between $110$ and $600$ meters, and users may be situated anywhere from $0$ to $150$ meters, as we consider the maximum obstacle height to be $150$ meters. The minimum building height is considered to be $30$ meters. Moreover, we assumed that there are $25$ high-rise buildings in the area. The distance between each two edge points of buildings is also considered equal to $10$ meters. The capacity of the \acrshort{fbss}' battery ($\zeta$) is considered $15,000$ mAh. In addition, we assume the maximum distance ($D$) that one \acrshort{fbs} can fly with this battery to be $15$ km.  We also consider the \acrshort{fbs}'s hovering energy consumption to be equal to $1,000$ joules \cite{fotouhi2019survey}. The battery assumed in this article for each \acrshort{fbs} is a $3$-cell type, and its output voltage ($Battery_v$) is assumed to be $11.1$ volts.
We also consider $45$ degrees to be the elevation angle of \acrshort{fbss}. We use the Merge cell method proposed in \cite{Rahimi2021} for the positioning model to find candidate points. We also used the Cplex solver to solve the positioning and trajectory mathematical models. We ran the simulation and solved the models many times; the results of the simulation are presented below.

\begin{footnotesize}
\begin{table}[t]
\caption{Test parameters to evaluate the problem model.}
\centering
\begin{tabular}{c|llll}
Parameters                      & \multicolumn{4}{c}{Description}                                       \\ \hline
$Region$                    &$5,000\times 5,000$ m                        \\
$U$                &$80$, $200$, $450$   \\
$\beta$                     &$100$ Mbps            \\
$D_{avg}$           &$3$ Mbps           \\    
$H_{{\rm min}}$ , $H_{{\rm max}}$          &$110$ m , $600$ m             \\  
$BH_{\rm min}$ , $BH_{\rm max}$          &$30$ m , $150$ m             \\ 
$B_{number}$                & $25$\\
Distance between edge points   & $10$ m\\
$f_c$               & $2$ GHz\\
{ $\mathsf{PL_{max}}$ }    & $110$ dB    \\
$a$                     & $9.61$\\
$b$                     & $0.16$\\
$\delta_{\mathsf{LoS}}$          & $1$\\
$\delta_{\mathsf{NLoS}}$         & $20$\\
$V_{\mathsf{FBS}}$               & $15$ m/s\\
$\zeta$                     & $15,000$ mAh\\
$D$                     & $15$ km\\
$Battery_{v}$               & $11.1$ v\\
$E_{\rm hover}$             & $1,000$ J

\end{tabular}
\label{testparameters}
\end{table}
\end{footnotesize}

\begin{figure}[b]
        \centering
        \includegraphics[width=0.9\columnwidth]{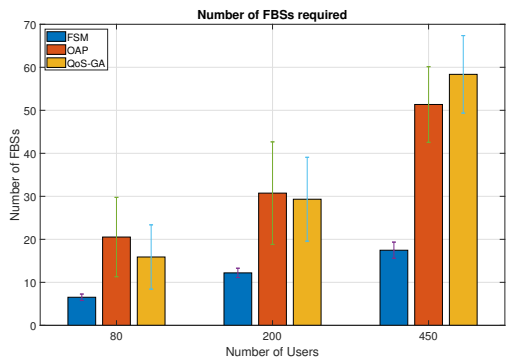}
        \caption{Comparison of FBSs required for different methods.}%
        {{\small }}    
        \label{FSMcomp}
\end{figure}

\subsection{Results}
In the following, we discuss the results of the proposed positioning and trajectory approach, compare the results of three different numbers of users. 
\par The number of required \acrshort{fbss} for each scenario with different methods is illustrated in Figure \ref{FSMcomp}. The proposed \acrshort{fsm} method is compared with the ordered artificial bee colony (ABC)-based placement (OAP) and QoS-GA methods. 
In the OAP algorithm, by grouping users into several groups, the service users of each UAV were first identified. One UAV serves each group of users. Then, each UAV's 3D position was chosen based on the clustering criteria \cite{zhang20213d}.
QoS-GA is a GA-based FBSs deployment technique which has been developed to maximize the number of covered UEs while also satisfying their data rate needs within the FBSs' capacity constraint. 
The population size, crossover rate, and mutation rate are three variables that affect the fundamental GA process. The number of potential solutions depends on the size of the population. The crossover rate and mutation rate values reflect the variety of potential solutions throughout an iteration. Up until the time step hits an iteration limit, the entire process is repeated.
The GA model treats each FBS's horizontal location and coverage radius as a gene. Some iterations are then run to determine the 2D deployment outcome \cite{zhong2020qos}.

As we can see, the proposed method needs fewer \acrshort{fbss} to cover all users in different scenarios.
In the proposed \acrshort{fsm} method, on average, for $80$ users, about $6$ \acrshort{fbss} are needed to cover all users in each snapshot. For $200$ and $450$ users, on average, about $12$ and $17$ \acrshort{fbss} are needed, respectively.

\begin{figure}[t]
        \centering
        \begin{subfigure}{\columnwidth}
            \centering
            \includegraphics[width=0.9\columnwidth]{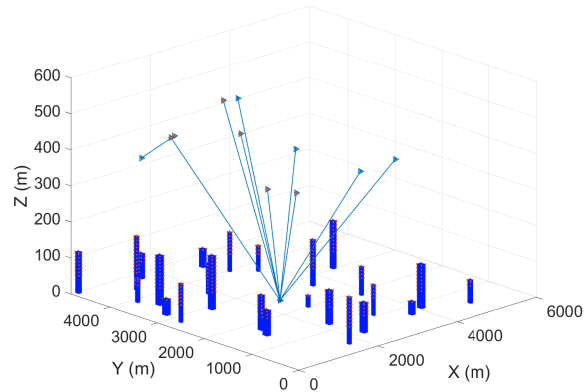}
            \caption{FBS trajectories phase $1$.}%
            {{\small }}    
            \label{tr1}
        \end{subfigure}
        \hfill
        \begin{subfigure}{\columnwidth}  
            \centering 
            \includegraphics[width=0.9\columnwidth]{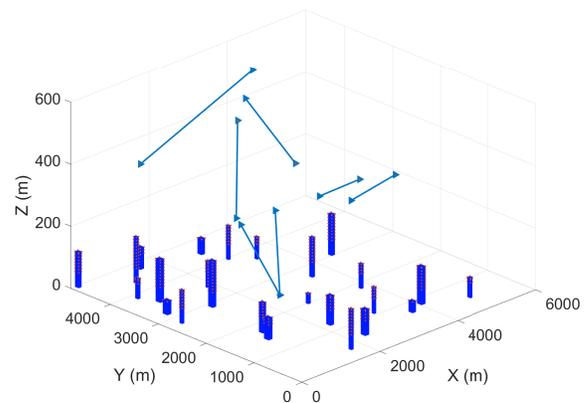}
            \caption{FBS trajectories phase $2$.}%
            {{\small }}    
            \label{tr2}
        \end{subfigure}
        \caption{FBSs trajectories in two subsequent snapshots in the $80$-user scenario.} 
        
        \label{trs}
\end{figure}

\begin{figure}[t]
        \centering
        \begin{subfigure}{\columnwidth}
            \centering
            \includegraphics[width=0.9\columnwidth]{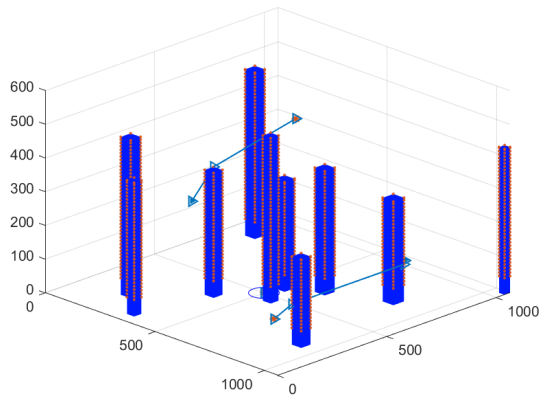}
            \caption{FBS trajectories and obstacle avoidance - 3D view.}%
            {{\small }}    
            \label{obstacles1}
        \end{subfigure}
        \hfill
        \begin{subfigure}{\columnwidth}  
            \centering 
            \includegraphics[width=0.9\columnwidth]{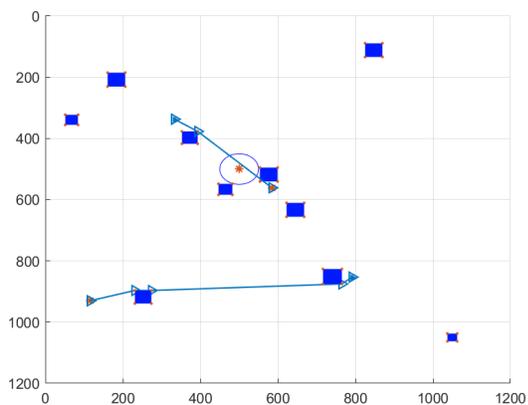}
            \caption{FBS trajectories and obstacle avoidance - 2D view.}%
            {{\small }}    
            \label{obstacles2}
        \end{subfigure}
        \caption{FBSs trajectories and their obstacle avoidance.} 
        
        \label{obstacles}
\end{figure}


\par Each snapshot is $15$ seconds away from the next one, and between each two positioning snapshots, a trajectory problem is solved. Figure \ref{trs} shows the \acrshort{fbss} trajectories in two different snapshots. In Figure \ref{tr1}, it is shown that \acrshort{fbss} fly from red points to blue points. Some of them increases its altitude while others does the opposite. This happens because some users move to different positions and a group of them may become very dense. Also, since the power of some of \acrshort{fbss} is not enough to continue their mission, the \acrshort{fbss} which were operated in the last snapshot returned to the base, and the two alternative full-power \acrshort{fbss} flew to the red points to operate. In addition, the two \acrshort{fbss} that returned to the base will be recharged for further missions. Moreover, as we need fewer \acrshort{fbss} than the previous snapshot, the extra FBSs that returned to the base will recharge and wait for future missions.
Figure \ref{tr2} illustrates that in this snapshot we need two more \acrshort{fbss} to cover users. Therefore, two new \acrshort{fbss} fly from the base to the destination points.
The power of the last two \acrshort{fbss} is not enough for the next operation. All of these trajectories are designed taking obstacles and collision avoidance into consideration.


\begin{figure}[t]
        \centering
        \includegraphics[width=0.9\columnwidth]{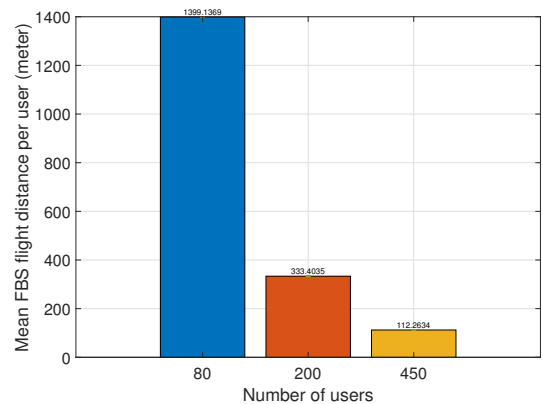}
        \caption{Mean \acrshort{fbs} flight distance per user.}%
        {{\small }}    
        \label{distance}
\end{figure}

\par A scenario of how two FBSs move inside the target area is presented 3D in Figure \ref{obstacles1} and 2D in Figure \ref{obstacles2}. Both Figures illustrate the trajectory within a snapshot from $t_1$ to $t_2$. FBSs move from their origin to the destination during the snapshot avoiding obstacles by applying the proposed algorithm. It is worth mentioning that the path used by FBSs is the shortest possible path between the origin and the destination which is presented in 2D and 3D. Red dot in \ref{obstacles} illustrates the place of the base where FBSs can recharge. 

\par Figure \ref{distance} presents a comparison of the mean FBS flight distance per user during the FBS operations. As the number of users increases, because we have more \acrshort{fbss}, the mean flight distance of an \acrshort{fbs} decreases. When we have fewer \acrshort{fbss}, the total distance of \acrshort{fbss} will increase.

\par
\begin{figure}[b]
        \centering
        \includegraphics[width=0.9\columnwidth]{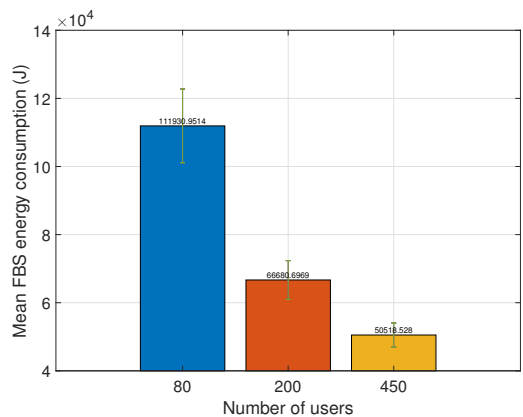}
        \caption{Mean \acrshort{fbs} energy consumption.}%
        {{\small }}    
        \label{energy}
\end{figure}

 In Figure \ref{energy}, the average \acrshort{fbs} flight energy consumption is compared in different scenarios. Energy includes the hovering and trajectory power consumed. As expected, with the increase in the number of users, the average power consumption of each \acrshort{fbs} must decrease because of shorter paths for \acrshort{fbss} and, therefore, less recharging. For example, with $80$ users, the number of \acrshort{fbss} is fewer than in other scenarios. As the mean flight distance of each \acrshort{fbs} is greater, the mean energy consumption of each \acrshort{fbs} is more significant than in other scenarios.

\begin{figure}[t]
        \centering
        \includegraphics[width=0.9\columnwidth]{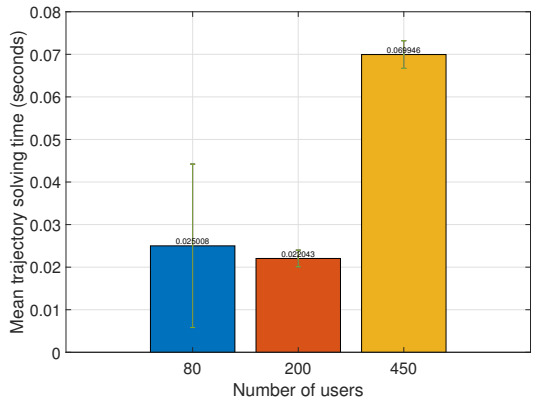}
        \caption{Mean trajectory solving time.}%
        {{\small }}    
        \label{time}
\end{figure}
\par Figure \ref{time} shows the comparison of the average solving time of the trajectory problem. As we can see, besides the proposed trajectory mathematical model having an exact solution, the time it takes to solve the problem is less than the time interval between two snapshots. It means we have enough time to solve the problem in the real world. By the way, we might add that a greater number of FBSs will mean more time to solve the problem. However, sometimes to have the optimum solution of the problem and minimize the \acrshort{fbss} energy consumption considering the constraints of the problem, the solving time can increase. As the figure shows, the average time it takes to solve the $80$-user scenario is somewhat longer than the $200$-user scenario because of some data calculations. It is worth noting that the standard deviation of trajectory solving time with $80$ users shows that in some cases, the solving time of this scenario is less than the minimum solving time of other scenarios.

\begin{figure}[b]
        \centering
        \includegraphics[width=0.9\columnwidth]{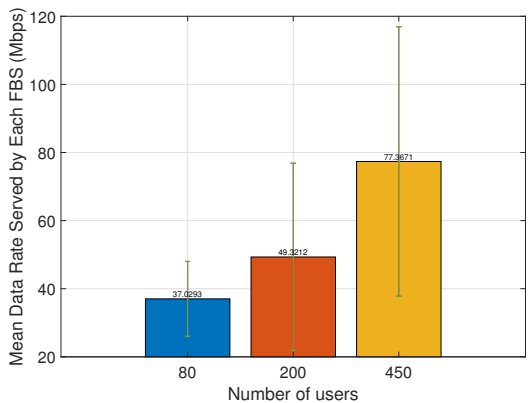}
        \caption{Mean data rate served by each FBS.}%
        {{\small }}    
        \label{data_rate}
\end{figure}

\begin{figure}[t]
        \centering
        \includegraphics[width=0.9\columnwidth]{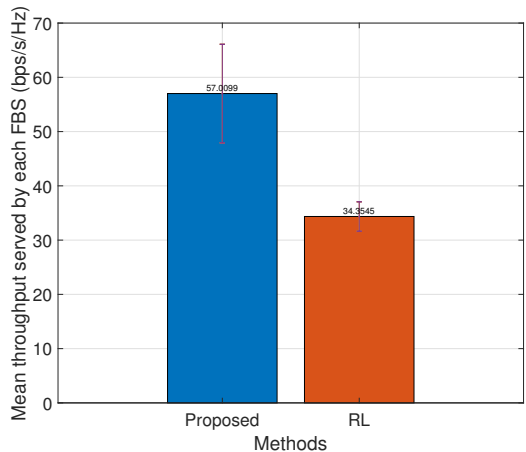}
        \caption{Comparison of throughput of RL \cite{9701330} and proposed method.}%
        {{\small }}    
        \label{CompWithRL}
\end{figure}

Figure \ref{data_rate} shows that with an increasing number of users, the mean data rate served by each \acrshort{fbs} increases. Therefore, we can conclude that the backhaul's efficiency will increase with a greater number of users, and more backhaul will be used in each small cell during the operation.

\par Figure \ref{CompWithRL} presents the throughput of the proposed algorithm compared with \cite{9701330}. The algorithm of \cite{9701330} is selected because it uses RL based algorithm for trajectory. We have proved the efficiency of the proposed algorithm in \ref{shortest_global}; this evaluation is used as a verification for the claim. Using the mean of achieved throughput in several runs, throughput of the proposed algorithm is ~$57$ and ~$34$ for \ref{shortest_global}. The minimum achieved throughput of the proposed algorithm in several runs is more than the maximum achieved throughput of \ref{shortest_global}. It is worth mentioning that \ref{shortest_global} uses only one FBS, does not consider obstacles, does not consider energy, and does not report its processing and train time. We have reported the energy consumption in \ref{energy} and processing time in \ref{time}. The proposed algorithm achieves better result despite it uses more real-world characteristics. There are 40 users in a

\section{Conclusion}\label{conclusion}
This paper proposed an approach with an exact solution for the problem of multiple $3$D trajectories for FBSs, while considering constraints related to energy consumption, operational time, flight distance, inter-cell interference, and collision avoidance.
The approach consisted of two phases, namely an FBS positioning phase and an \acrshort{fbs} trajectory phase, where we divided the problem into several snapshots. First, we found the minimum number of FBSs required along with their optimal $3$D positions in each snapshot. Then, the trajectory phase ran between every two snapshots. The optimum path between the origin and the destination in the trajectory phase was found using the proposed binary linear problem (BLP) model, considering \acrshort{fbs} energy consumption and flight distance limitations. As the proposed method was a BLP, the solution was the optimal one. We used a shortest path heuristic to find the best path of each \acrshort{fbs} from origin to the destination, taking the constraint of collision avoidance into account. In different snapshots, the required number of \acrshort{fbss} could be different. To address this, we introduced the FSM approach to manage the set of \acrshort{fbss} and their power. The results showed that the proposed method is operational in real-world scenarios. The results also showed that the mean \acrshort{fbs} flight distance per user and mean \acrshort{fbs} energy consumption decreased as the number of users increased. Also, the mean data rate served by each \acrshort{fbs} was shown to increase as the number of users increased.

\appendix
\label{appendix1}
The proof for the global and definite solution of the proposed algorithm
\begin{theorem}{Suppose there is no direct path between points A and B (the connecting line between A and B passes through an obstacle), then the path obtained from the proposed algorithm based on the proposed graph and Dijkstra's algorithm will be the shortest path from A to B.}
\label{shortest_global}
\end{theorem}
\begin{proof}
There are two cases for the problem, 1) where the path only passes one vertex of the obstacle, and 2) where the path  passes two vertexes of the obstacle. We will provide a proof for the theorem for each case in the following. 

1) The path only passes one vertex of the obstacle, from point A to M and from M to B.

This case includes two sub-cases which will be presented in 1-a and 1-b. 
1-a) In this sub-case, as you can see in Figure \ref{d1}, if a curved path from point A (source) to B (destination) exists, the shortest path would be a path that passes point M (vertex of the obstacle).

\begin{figure}[h]
        \centering
 \includegraphics[width=0.9\columnwidth]{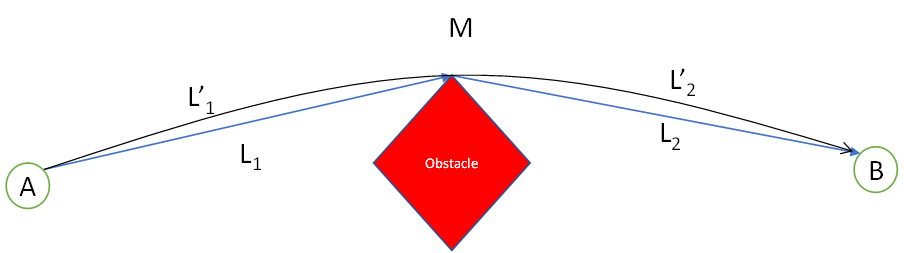}
        \caption{Proof for case 1-a, where there is curved line, presenting point M.}%
        {{\small }}    
        \label{d1}
\end{figure}
To illustrate the proof, we assume that there is another curved path from A to B that passes a point N as shown in Figure \ref{d2}.
\begin{figure}[h]
        \centering
 \includegraphics[width=0.9\columnwidth]{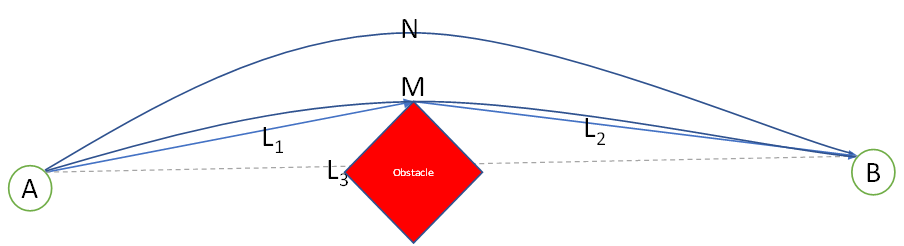}
        \caption{Proof for case 1-a, where there is curved line, presenting point N.}%
        {{\small }}    
        \label{d2}
\end{figure}

We consider the line between A and B as the axis of a coordinate plane, so the area under the curved line passes point N is more than the area under the curved line passes point M. The area under the curved line $L$ is calculated as follows:

\begin{equation}\label{proof_1}
    \mathsf{L} = \int_{A}^{B}\sqrt{1+[f'(x)]^2}dx.
\end{equation}
        
Since point M is closer to the axis than N (the curved line can not pass through the obstacle), we have $\small\int_{A}^{B}f_{M}(x)dx < \int_{A}^{B}f_{N}(x)dx$. Therefore the shortest curved line passes point M. 
Noting that the curved line from A to M $(L'_1)$ is longer than the straight line from A to M $(L_1)$ and the curved line from M to B $(L'_2)$ is longer than the straight line from M to B $(L_2)$, therefore, the curved line from A to B ($L'_1+L'_2$) is longer than the straight line from A to M and then from M to B ($L_1+L_2$). Since the point N is an arbitrary point, we can reason that every other path from A to B is longer than the path that passes M. 
        
1-b) In the other sub-case, as shown in Figure \ref{d3}, we assume that there is a straight line from A to N $(L'_1)$ and then from N to B $(L'_2)$.

\begin{figure}[h]
        \centering
 \includegraphics[width=0.9\columnwidth]{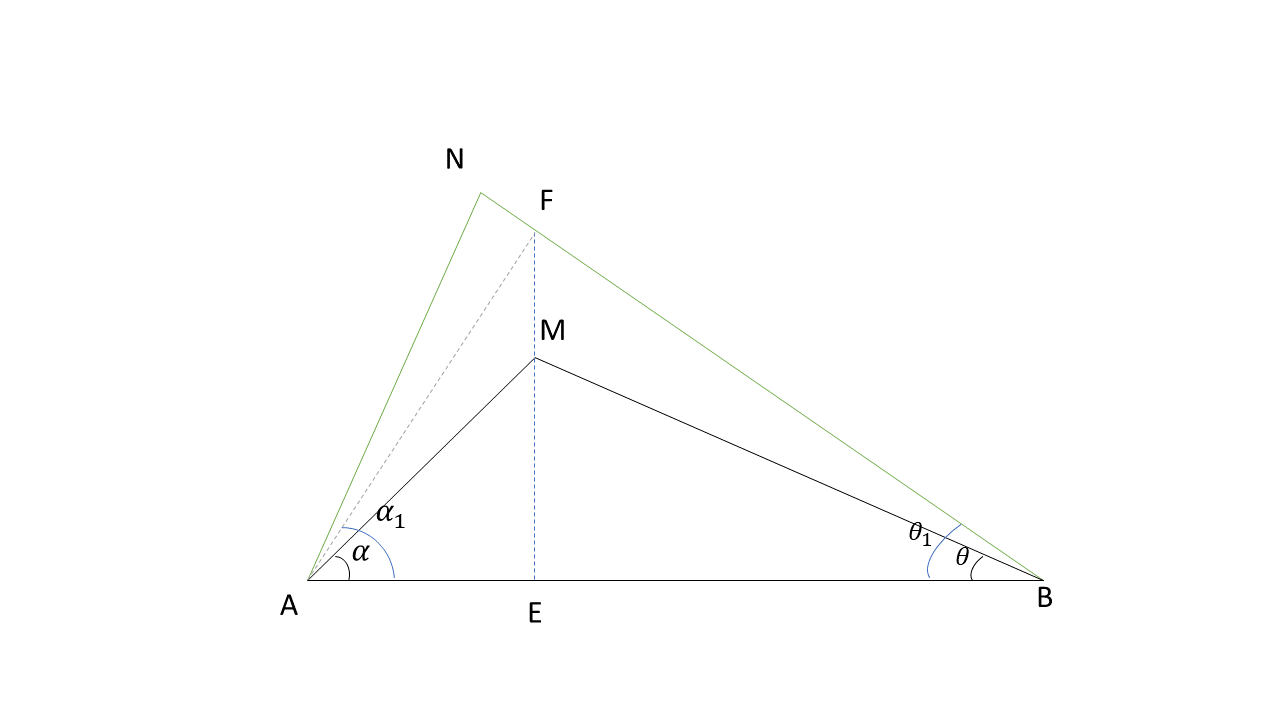}
        \caption{Proof for case 1-b, where there is straight line, presenting point M and N.}%
        {{\small }}    
        \label{d3}
\end{figure}

where $ \theta<\theta_1 \rightarrow \cos{\theta}>\cos{\theta_1} \rightarrow \frac{BE}{BM}>\frac{BE}{BF} \rightarrow BF>BM$.

Now, to achieve the goal, we have to prove that $FN+NA>AM$. According to \textit{"Triangle Sides Inequality Theorem"} we have $AF<AN+NF$. Similar to the fact $\theta<\theta_1$ we have $\alpha<\alpha_1$, which leads to know $AF>AM$. 
As a result we will have:

\begin{equation}\label{proof_2}
    (AF<AN+NF , AF>AM) \rightarrow AN+NF>AM.
\end{equation}

1-b') The sub-case 1-b, can also be proved in the following way:

As shown in \ref{d4}, since the bases of the two triangles T and T' are equal (the straight line from A to B (L\textsubscript{3})), and also considering that no path from A to the hypothetical point N and from N to B can cross the obstacle, it can be concluded that h<h', as a result, the area of triangle T is less than the area of triangle T'.

\begin{figure}[h]
      \centering
        \includegraphics[width=0.9\columnwidth]{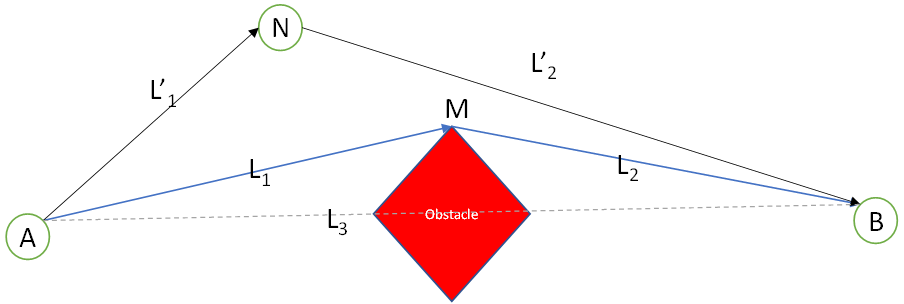}
        \caption{Another proof for case 1-b.}%
        {{\small }}    
        \label{d4}
\end{figure}

According to these assumptions, based on Heron's theorem, it can be proved that for T triangle
\begin{equation}\label{proof_3}
    A_T = \sqrt{S_T (S_T - L_1) (S_T - L_2) (S_T - L_3)},
\end{equation}
where $S_T = \frac{L_1+L_2+L_3}{2}$. Also for T' triangle 
\begin{equation}\label{proof_4}
    A_T' = \sqrt{S_T' (S_T' - L'_1) (S_T' - L'_2) (S_T' - L_3)},
\end{equation}
where $S_T' = \frac{L'_1+L'_2+L_3}{2}$.

According to $A_T$ and $S_T$, the $A_T$ can be obtained as follows:
\begin{equation}\label{proof_5}
    A_T = 0.25*\sqrt{(L^2_1+L^2_3+L^2_3)^2 -2(L^4_1L^4_+L^4_2+L^4_3)}.
\end{equation}

Since $A_T < A_T'$, so we need to prove the correctness of \eqref{proof_6} and \eqref{proof_7}.
\begin{equation}\label{proof_6}
    (L^2_1+L^2_3+L^2_3)^2 < (L'^2_1+L'^2_3+L^2_3)^2,
\end{equation}
\begin{equation}\label{proof_7}
    -2(L^4_1+L^4_2+L^4_3) > -2(L'^4_1+L'^4_2+L^4_3).
\end{equation}

In \eqref{proof_6}, considering that $L_3$ is common and equal in both triangles, as a result $(L^2_1+L^2_2)^2 < (L'^2_1+L'^2_2)^2$. So $L^2_1+L^2_2 < L'^2_1+L'^2_2$. Since the length of the triangle is always a positive value,
\begin{equation}\label{proof_8}
    L_1+L_2 < L'_1+L'_2.
\end{equation}

We can draw conclusions for \eqref{proof_7} as well,
\begin{equation}\label{proof_9}
    (L^4_1+L^4_2+L^4_3) < (L'^4_1+L'^4_2+L^4_3).
\end{equation}

Since $L_3$ is common to both triangles and the length of the sides is always positive, we can also conclude from \eqref{proof_9} that $L_1+L_2 < L'_1+L'_2$. As a result, the path passing through the edge of the obstacle is the shortest possible path from A to B.
\\

2) The path passes two vertexes of the obstacle, from point A to M\textsubscript{1}, from M\textsubscript{1} to M\textsubscript{2} and then from M\textsubscript{2} to B.

In this case, it will be proved for each of the triangles $AM_1M_2$ and $M_1M_2B$ that they present the shortest path, likewise the proof 1. Now, we have to prove that for any arbitrary point N, the path passing N, will be longer than path mentioned in this case passing two vertexes.

\begin{figure}[h]
        \centering
 \includegraphics[width=0.9\columnwidth]{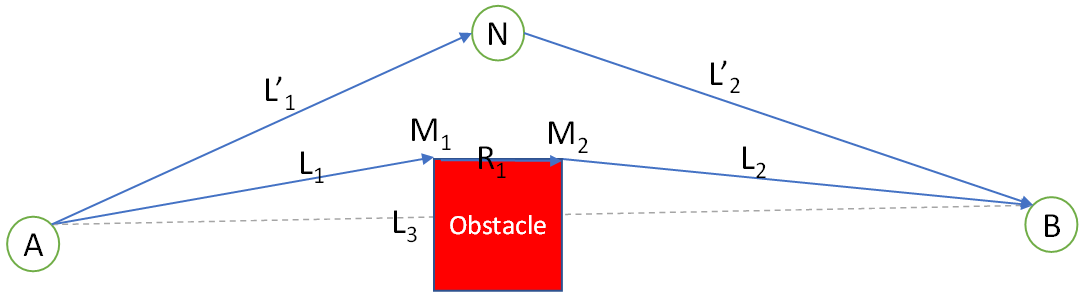}
         \caption{Proof for case 2, where there is straight line, presenting point N.}%
        {{\small }}    
        \label{d6}
\end{figure}

As presented in \ref{d6}, knowing the fact that straight line from point A to M\textsubscript{1} is shortest path from A to the vertex of the obstacle (M\textsubscript{1}), if the length of the straight line from A to N is shorter than the straight line from A to M\textsubscript{1}, or similarly, if the length of the straight line from N to B is shorter than the straight line from M\textsubscript{2} to B, the path passing N hits the obstacle or it is tangent to either the straight line from A to M\textsubscript{1} or the straight line from M\textsubscript{2} to B as presented in \ref{d7}.

\begin{figure}[h]
        \centering
 \includegraphics[width=0.9\columnwidth]{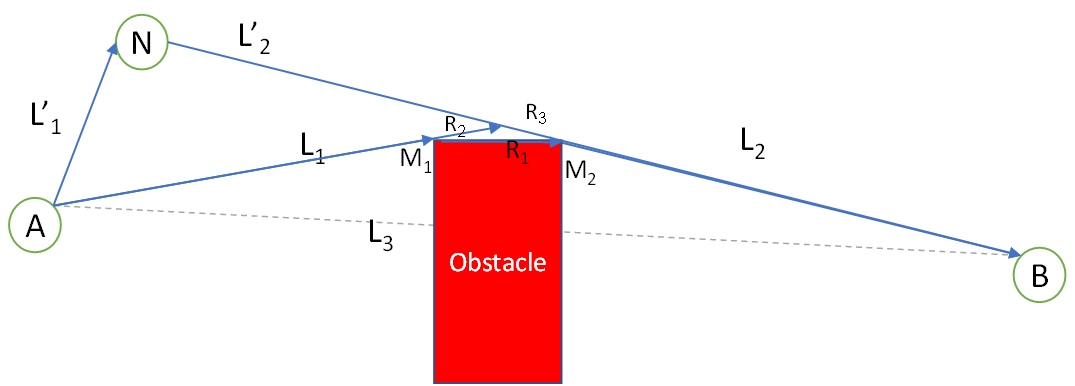}
        \caption{Proof for case 2, presenting the contradiction of the path passing point N.}%
        {{\small }}    
        \label{d7}
\end{figure}

If one of the paths is tangent, according to \textit{"Triangle Sides Inequality Theorem"}, $L_1+R_2<L'_1+L'_2$ and since $R_1<R_2+R_3$ then $R_1-R_2,R_3$. Therefore $L_1+R_1<L'_1+L'_2+R_3$.
Now, it is concluded that the length of every path from A to N that does not pass M\textsubscript{1} is longer than the path the passes M\textsubscript{1} based on the \textit{"Triangle Sides Inequality Theorem"}. Similarly, the path from N to B must pass M\textsubscript{2} as depicted in \ref{d9}.

\begin{figure}[h]
        \centering
 \includegraphics[width=0.9\columnwidth]{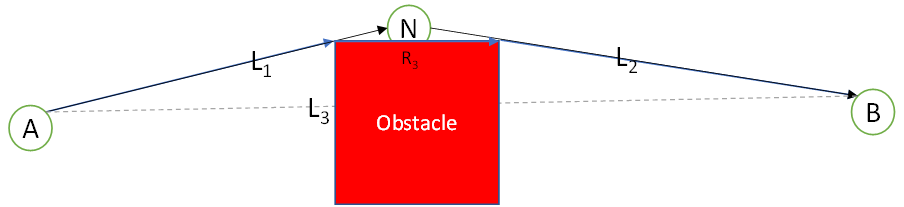}
        \caption{Proof for case 2, presenting the contradiction of the path passing point N where one or ore lines are tangent.}%
        {{\small }}    
        \label{d9}
\end{figure}

Knowing the fact that the path including A, N, B must be rewritten as A, M\textsubscript{1}, N, M\textsubscript{2}, B; according to the \textit{"Triangle Sides Inequality Theorem"} the A, N, B path could not be shorter than  A, M\textsubscript{1}, M\textsubscript{2}, B path. Therefore, the path that passes the vertexes is the shortest.
\end{proof}

\bibliographystyle{IEEEtran}
\bibliography{main.bib}
\end{document}